\documentclass[aps,prx,reprint,longbibliography,showkeys]{revtex4-2}
\usepackage{amsmath,amssymb}
\usepackage{graphicx}

\begin{document}

\title{Operational Relation Between One-Time and Two-Time Work Protocols and Measurement-Resolution-Induced Transitions in Quantum Work Statistics}

\author{Daniel Alonso and Antonia Ruiz-García}
\affiliation{Departamento de F\'\i sica and IUdEA, Universidad de La Laguna}
\email{dalonso@ull.edu.es}
\date{\today}

\begin{abstract}
	
We establish a direct operational connection between the one-time measurement (OTM) and two-time measurement (TTM)
protocols for quantum work statistics, showing that OTM work values, including coherence signatures, can be
reconstructed from standard TTM data through classical post-processing without any modification of the experimental
setup. This reveals that coherence information commonly attributed to OTM schemes is already latent within TTM
data and becomes explicit upon a natural post-processing step. For a driven two-level system subject to
finite-resolution energy measurements, this reconstruction unveils a resolution-driven statistical transition in the
work distribution: whereas projective and weak measurements yield smooth distributions, intermediate resolutions produce
non-analytic structures whose character is determined by the dynamics and a critical meter
resolution. We show that the non-analyticity of the OTM work distribution takes the form of square-root divergent peaks
at specific work values, and that the separation between these singular points is an experimentally accessible
quantity that vanishes at the critical resolution with a universal exponent. The same dynamical quantities that govern the transition also encode the relative entropy of coherence of the driven state, enabling coherence quantification from energy
measurements alone. We demonstrate experimental accessibility in a nitrogen-vacancy center platform with parameters drawn from current experiments, where both sides of the transition and the coherence signatures are within reach of existing technology.

\end{abstract}

\maketitle

\section{Introduction}

Fluctuation theorems are central to our understanding of systems far from equilibrium \cite{evans1993probability,gallavotti1995dynamical,gallavotti1995dynamicalensembles,jarzynski1997nonequilibrium,crooks1999a}. They generalize the second law of thermodynamics \cite{jarzynski2011equalities} and earlier results in statistical physics, such as linear response theory \cite{andrieux20081,andrieux2009thefluctuation,gaspard2022statistical}, and have been experimentally verified in quantum setups including NMR \cite{batalhao2014experimental} and nitrogen-vacancy (NV) centers in diamond \cite{hernandez2020experimental}.

In quantum thermodynamics, a standard operational definition of work is based on the two-time measurement (TTM) protocol. Projective energy measurements are performed before and after a driving protocol generated by a time-dependent Hamiltonian $\hat H(t)$, and work is defined as the difference of the two outcomes \cite{jarzynski2011equalities,esposito2009nonequilibrium,campisi20111,kurchan2000aquantum1,tasaki2000jarzynski}. The measurement process destroys coherences in the energy basis, in particular, because the second measurement destroys coherence in the final energy basis, TTM is often regarded as insensitive to quantum coherences that may develop during the dynamics after the first energy measurement.

One-time measurement (OTM) schemes have been proposed as an alternative. In OTM, work is defined as the difference between the initial energy measurement and the expectation value of the energy at the final time \cite{deffner2016quantum,sone2021jarzynski}. These protocols have been related to nondemolition measurements and quantum circuit implementations and extended to open systems \cite{maeda2023detailed,sone2020quantum,oftelie2025measurement}. They are frequently interpreted as more sensitive to coherence than TTM.

At the same time, a variety of other approaches to quantum work have been explored, such as interferometric schemes using ancillas \cite{roncaglia2014work,chiara2018ancilla}, collective measurements on many copies of the system \cite{acin2017nogo}, and trajectory-based approaches that assign work and heat along stochastic quantum trajectories \cite{seifert2012stochastic,elouard2017therole}. A development of work fluctuation relations for \emph{generalized} measurements is found in \cite{watanabe2014generalized, ito2019generalized}. Another approach consist in considering a quantum particle and its scattering with the system of interest. At this respect, it has been shown that the energy fluctuations obtained within the standard TTM protocol can be accessed in an alternative, fully autonomous setting, where a quantum particle scatters off the system of interest. In this scenario, the incoming particle simultaneously acts as the external drive and as an energy meter for the system, so that the corresponding work statistics can be reconstructed from the particle’s kinetic energy distribution \cite{jacob2023twopoint,jacob2024universal}.

More recently, the role of finite measurement resolution and meter quantum fluctuations has been emphasized in a series of works \cite{alonso2023single,han2024quantum,varma2025quantum} which highlight fundamental limitations and new possibilities in quantum thermodynamic experiments. 

Despite this progress, the precise relationship between TTM and OTM remains subtle. Do they contain fundamentally different information? It is usually stated that OTM is more sensitive to coherence. Do standard TTM data already contain coherence information that is simply not extracted in usual analyses? How does finite measurement resolution alter this picture?

In this work, we address these questions and emphasize three main results:

First we show that data obtained via the standard TTM protocol can be classically post-processed to reconstruct the corresponding OTM work values. Thus, OTM can be viewed as an operational refinement of TTM, in the sense that conditional work values can be reconstructed from standard TTM data through classical post-processing. In particular, we derive an explicit relation expressing OTM work as a conditional average of TTM work over the second measurement outcome. This leads to the extraction of coherence from the TTM protocol data.

Second, for a driven two-level system subject to finite-resolution energy measurements, we identify a resolution-induced non-analytic change in the structure of the work distribution, which we refer to as a statistical transition, governed by the sign of a dynamical quantity $A_1(t)$ and a critical meter resolution $\sigma_c$. In the limits of projective or very weak measurements,
the OTM work distribution is smooth (Gaussian or a sum of Gaussians). At intermediate resolution, however, non-analytic,
peak-like structures emerge when the OTM work $W(t,f)$ becomes multivalued as a function of the first outcome $f$. We 
provide a quantitative characterization of this transition: the separation between the two singular work values vanishes at
$\sigma_c$ as $\Delta W \sim \varepsilon^{3/2}$, where $\varepsilon = (\sigma_c-\sigma)/\sigma_c$, a universal exponent, independent on the Hamiltonian or populations. Furthermore, the work distribution itself diverges at the singular work values as $P_W(w) \sim (\sigma_c-\sigma)^{-1/4}|w-w_i|^{-1/2}$, providing two independently measurable signatures of the critical point.
	
Third, we show how, for two-level systems, the OTM work distribution, which can be obtained from the TTM data, contains enough information to reconstruct a standard measure of quantum coherence, namely the relative entropy of coherence in the instantaneous energy basis at the final time. This provides an experimentally feasible method for accessing coherence using only energy measurements. While this reconstruction is demonstrated explicitly for two-level systems, it illustrates how work statistics can encode coherence information in experimentally relevant settings.

We illustrate our findings in an NV-center platform modeled after \cite{hernandez2020experimental}, and demonstrate that the predicted statistical transition and coherence signatures are accessible with current technology.

\section{Relation between TTM and OTM}\label{TTMOTM}

Given a system with Hamiltonian $\hat H(t)$, we model nonideal energy measurements at time $t$ by a family of Positive Operator Valued Measurments (POVM),
\begin{align}
	\hat G_{\sigma}^{1/2}(F|\hat H(t)),
\end{align}
where $F$ denotes the measurement outcome with realizations $f$ and resolution $\sigma$. The identity can be expressed as 
\begin{align}
	\hat 1=\int df \hat G_{\sigma}^{1/2}(f|\hat H(t)) \hat G_{\sigma}^{1/2 \dagger}(f|\hat H(t)).
\end{align}
Writing the spectral decomposition
\begin{align}\label{Hspectral}
	\hat H(t)=\sum_{i=1}^N \mu_i(t) |\mu_i(t)\rangle\langle \mu_i(t)|,
\end{align}
we consider POVM elements of the form
\begin{align}
	\hat G_{\sigma}^{1/2}(F|\hat H(t)) = \sum_{i=1}^N G_{\sigma}^{1/2}(F|\mu_i(t)) |\mu_i(t)\rangle\langle \mu_i(t)|,
\end{align}
with $G_{\sigma}^{1/2}(F|\mu_i(t))$ real, positive, and localized around $\mu_i(t)$ with resolution $\sigma$. In the limit $\sigma\to 0$, these operators approach projectors onto the energy eigenstates and the measurement becomes projective.

If we start from an initial state $\rho(0)$, then, after an energy measurement at $t=0$ with outcome $f_1$, the conditional post-measurement state is given by
\begin{align}
	\rho(0^+|f_1) = \frac{\hat G_{\sigma}^{1/2}(f_1|\hat H(0)) \rho(0) \hat G_{\sigma}^{1/2}(f_1|\hat H(0))}{P_F(f_1)},
\end{align}
with $P_F(f_1)=\mathrm{Tr}\,\left[ G_{\sigma}(f_1|\hat H(0)) \rho(0)\right]$, the probability density of the outcome $f_1$. If $\hat U(t)$ is the unitary evolution operator associated to $\hat H(t)$ we can write for the conditional state at time $t$
\begin{align}
	\rho(t|f_1)=\hat U(t)\rho(0^+|f_1),\hat U^{\dagger}(t).
\end{align}

In the TTM protocol, a second energy measurement at time $t$ yields outcome $f_2$ and a post-measurement state
\begin{align}
	\rho(t^+|f_2,f_1)=\frac{\hat G_{\sigma}^{1/2}(f_2|\hat H(t)) \rho(t|f_1) \hat G_{\sigma}^{1/2}(f_2|\hat H(t))}{P_F(f_2|f_1)},
\end{align}
where
\begin{align}
	P_F(f_2|f_1)=\mathrm{Tr}\left[\hat G_{\sigma}(f_2|\hat H(t)) \rho(t|f_1)\right]
\end{align}
is the conditional probability of observing $f_2$ given $f_1$ \cite{braginsky1995quantum}. Work is defined as
\begin{align}
	W^{(2)}(f_2,f_1) = f_2-f_1.
\end{align}

In the OTM protocol, the second measurement is replaced by the expectation value of energy at time $t$, conditioned on the first outcome $f_1$ \cite{deffner2016quantum}:
\begin{align}
	W^{(1)}(f_1,t) = \mathrm{Tr}\left[\hat H(t)\rho(t|f_1)\right]-f_1.
\end{align}

In fact, the two schemes are directly related. Averaging the TTM definition of work over $f_2$ with the conditional probability $P_F(f_2|f_1)$ yields
\begin{align}
	\int df_2  W^{(2)}(f_2,f_1) P_F(f_2|f_1)
	&= \int df_2 \,(f_2-f_1) \,P_F(f_2|f_1)\nonumber \\
	&= \mathrm{Tr}\left[\hat H(t)\rho(t|f_1)\right]-f_1 \nonumber \\
	&= W^{(1)}(f_1,t),
\end{align}
where we have used the spectral representation of $\hat H(t)$ and the property $\int df \, f \, G_{\sigma}(f|\mu)=\mu$ for each eigenvalue $\mu$. Thus,
\begin{equation}\label{conditional}
	W^{(1)}(f_1,t)=\int df_2 \, W^{(2)}(f_2,f_1)\,P_F(f_2|f_1),
\end{equation}
showing that \emph{OTM work values can be reconstructed directly from TTM data} via classical post-processing of the second measurement outcomes. A sufficiently large TTM data set yields the conditional probabilities $P_F(f_2|f_1)$ required for this reconstruction. A classical analogue of Eq. (\ref{conditional}) was discussed in \cite{sone2021jarzynski}, where $W^{(1)}(f_1,t)$ was termed conditional work. Our contribution advances this picture in two respects. First, we work within a fully quantum, non-projective measurement framework that enables coherence effects to be tracked at finite resolution $\sigma$. Second, Eq. (\ref{conditional}) establishes operationally that a standard TTM experimental run without any modification to the measurement apparatus already accumulates sufficient statistics to reconstruct the OTM work values and, consequently, coherence-sensitive quantities. In this sense, coherence information is latent within the TTM data, and the OTM distribution arises as a natural post processing step that renders such coherences explicit. Finally, as we demonstrate below for the two-level system, the entropy of coherence is directly computable from the OTM data. Taken together, these results provide a new perspective on the relationship between the TTM and OTM schemes and on the role of coherences in work statistics. 

We note that while Ref.~[30] established the generalized Jarzynski relation for the OTM protocol within a single-measurement framework and generalized measurements, the present work derives the explicit operational equivalence between TTM and OTM, identifies and quantitatively characterizes the resolution-driven statistical transition in the work distribution, and demonstrates how the resulting OTM data encode the relative entropy of coherence of the driven state.

\section{OTM work distribution and generalized Jarzynski relation}\label{OTMdist}

From now on we denote $W^{(1)}(t,f)\equiv W(t,f)$. Its fluctuations are characterized by
\begin{eqnarray}
	P_W(w)=\int df\, P_F(f)\, \delta \bigl(w -W(t,f)\bigr),
\end{eqnarray}
which captures the randomness due to measurements.

The average work is
\begin{align*}
	\langle W(t,f)\rangle &\equiv \int df\, P_F(f) W(t,f) \nonumber \\ 
	&={\text Tr}\big[\hat H(t) \rho(t)\big]-{\text Tr}\big[ \hat H(0) \rho(0)\big],
\end{align*}
with $\rho(t)=\hat U(t)\rho(0)\hat U^{\dagger}(t)$, in agreement with the first law of thermodynamics \cite{kosloff2013quantumthermodynamics,deffner2019quantumthermodynamics}.

For a reference thermal state $\rho_{th}(t)=e^{-\beta \hat H(t)}/{{\cal{Z}}(t)}=e^{-\beta (\hat H(t)-{\textsf{F}}(t))}$, with inverse temperature $\beta=1/k_B T$ and Helmholtz free energy $\textsf{F}(t)$, the OTM statistics satisfies a generalized Jarzynski relation \cite{deffner2016quantum,alonso2023single},
\begin{eqnarray}
	\left \langle e^{\beta (\Delta \textsf{F}-W)}\right \rangle=e^{\xi},
\end{eqnarray}
where $\Delta\textsf{F} = \textsf{F}(t) - \textsf{F}(0)$ and
\begin{eqnarray}
	\xi=\ln \left\langle e^{\beta ( f-\langle \hat  H(0) \rangle)} e^{-\Delta S} e^{-\Delta C}e^{-\Delta D_{KL}} \right\rangle.
\end{eqnarray}
Here $\Delta S = S(\rho(t)) - S(\rho(0))$ is the change in von Neumann entropy $S(\rho) = -\text{Tr}(\rho \ln \rho)$,
\begin{equation}\label{CDKL}
	\Delta C=C_{\hat H(t)}(\hat \rho(t,f))-C_{\hat H(0)}(\hat \rho(0))
\end{equation}
is the change in the relative entropy of coherence, defined as $C_{\hat H(t)}(\hat \rho(t,f))=S(\rho_D(t))-S(\rho(t))$ \cite{streltsov2017quantum}, and
\begin{equation}\label{DKL}
	\Delta D_{KL}=D_{KL}(\hat \rho_D(t,f)||\hat \rho_{th}(t))-D_{KL}(\hat \rho_D(0)||\hat \rho_{th}(0))
\end{equation}
gives the change in the Kullback–Leibler divergence between the decohered state $\rho_D(t)$ (diagonal in the energy basis) and the thermal state $\rho_{\text{th}}(t)$, with $D_{\text{KL}}(X \| Y) = \text{Tr}(X \ln X - X \ln Y)$.

Equation \eqref{CDKL} makes explicit how coherence enters the OTM fluctuation relation, while Eq. \eqref{DKL} quantifies the distance to the corresponding equilibrium state. A detailed discussion of these contributions and their dependence on measurement resolution can be found in \cite{alonso2023single}. From now on we focus on the structure of $P_W(w)$ in a two-level system.

\section{Two-level systems and structure of the work distribution}\label{TLS}

We focus on the archetypal case of a two-level system, where the analysis can be carried out in full analytical detail. We note, however, that the mechanism underlying the statistical transition, namely, the emergence of multiple branches in the conditional work function $W(t,f)$ when it becomes non-monotonic, is not specific to two-level systems. In a multilevel system with $N$ energy levels, the conditional work function involves a ratio of sums of $N$ Gaussian terms, which generically develops additional extrema at intermediate measurement resolutions. While a complete characterization of the critical structure in the general case is more complex, the essential ingredients  as finite-resolution POVM elements and non-trivial unitary mixing of populations, are present for arbitrary $N$. 

The initial state is taken diagonal in the eigenbasis of  $\hat H(0)$ with populations $p_1$ and $p_2$. Such state can be written as a thermal state with inverse temperature $\beta =\ln(p_1/p_2)/2\mu$, which turns out to be a convenient parameter in the following. Notice that in the case of population inversion the temperature can be negative. We shall thus write for the initial state
\begin{eqnarray}\label{inistate}
	\rho(0) &=& p(\mu_1(0),\beta)|\mu_1(0)\rangle\langle\mu_1(0)| \\ \nonumber &+&p(\mu_2(0),\beta)|\mu_2(0)\rangle\langle\mu_2(0)|,
\end{eqnarray}
with $p(\mu_i(0),\beta)=e^{-\beta \mu_i(0)}/(e^{-\beta \mu_1(0)}+e^{-\beta \mu_2(0)})$, $i \in \{1,2\}$.
It follows that the OTM work at time $t$ and outcome $f$ is
\begin{widetext}
	\begin{align}\label{eq:work}
		W(t,f) &= \mathrm{Tr}[\hat H(t) \hat U(t) \rho(0^+|f) \hat U^\dagger(t)] - f  \\
		&= \langle \mu_1(0) | \hat U^\dagger(t) \hat H(t) \hat U(t) | \mu_1(0) \rangle
		\frac{p(\mu_1(0),\beta) G_\sigma(f|\mu_1(0)) - p(\mu_2(0),\beta) G_\sigma(f|\mu_2(0))}
		{p(\mu_1(0),\beta) G_\sigma(f|\mu_1(0)) + p(\mu_2(0),\beta) G_\sigma(f|\mu_2(0))} - f,
	\end{align}
\end{widetext}
where
\begin{equation}
	\langle \mu_1(0) | \hat U^\dagger(t) \hat H(t) \hat U(t) | \mu_1(0) \rangle \equiv A_1(t)\in \mathbb{R}
\end{equation}
collects how $W(t,f)$ depends on the dynamics. 

In what follows, we model the meter by a Gaussian pointer,
\begin{equation}
	G_\sigma(f|\mu) = \frac{1}{\sqrt{2\pi \sigma^2}} \exp\Big[-\frac{(f-\mu)^2}{2\sigma^2}\Big].
\end{equation}
For this choice, taking into account that the Hamiltonian is traceless, then $\mu_2 =- \mu_1 = \mu$, Eq. \eqref{eq:work} can be written as
\begin{align}\label{eq:w_tanh_short}
	W(t,f) = A_1(t) \tanh\Big(\beta \mu - \frac{\mu f}{\sigma^2}\Big) - f.
\end{align}

The behavior of $P_W(w)$ as a function of measurement resolution is captured by the dimensionless parameter
\begin{equation}
	\lambda = \frac{2|\mu|}{\sigma},
\end{equation}
and by the sign of $A_1(t)$.

\subsection{Weak-measurement regime}

When the meter cannot resolve the two energy levels, $G_\sigma(f|-\mu) \approx G_\sigma(f|\mu)$, and $W(t,f)$ simplifies to
\begin{align}
	W(t,f) \simeq A_1(t) \Delta p - f,
\end{align}
with
\begin{align}
	\Delta p = p(-\mu,\beta) - p(\mu,\beta).
\end{align}
In this regime, $P_W(w)$ is essentially $P_F(f)$ with shifted argument
\begin{align}
	P_W(w) \simeq P_F(A_1(t) \Delta p - w),
\end{align}
with first and second moments as $\sigma \to \infty$ given by
\begin{align}
	\langle w \rangle &= (A_1(t) + \mu) \Delta p,  \nonumber \\
	\mathrm{Var}(w) &= \sigma^2.
\end{align}
A Gaussian profile is recovered in the limit of large $\sigma$. Although individual measurements carry little energy resolution, the ensemble statistics still allows extraction of $A_1(t)$ and $A_1(t)\Delta p$.

\subsection{Projective-measurement regime}

In the opposite case of projective measurements, each realization collapses the system onto one of the two eigenstates of $\hat H(0)$, leading to
\begin{align}
	W(t,f) =
	\begin{cases}
		-A_1(t) - f,  &\forall f < 0,\\
		A_1(t) - f,  &\forall f > 0.
	\end{cases}
\end{align}
In this case the work distribution is a superposition of two Gaussians, $i.e.$
\begin{align}
	P_W(w) = \mathcal{N} \big[G_\sigma(w|A_1(t) + \mu) + G_\sigma(w|-A_1(t) - \mu)\big],
\end{align}
where $\mathcal{N}$ normalizes the distribution. In this case, the meter resolution is sufficient to distinguish the energy eigenvalues, and the two peaks reflect the two possible energy transitions.

\subsection{Intermediate resolution}

An interesting situation arises at intermediate resolution for which $\sigma \approx 2 \mu$. In this case the function $W(t,f)$ becomes non-monotonic and develops multiple inverse branches $B$, $f = W_B^{-1}(w)$. The work distribution can then be written as
\begin{align}\label{eq:PWbranches}
	P_W(w) = \sum_{B} \frac{P_F(W_B^{-1}(w))}{\left|\frac{dW(f)}{df}\right|_{f=W_B^{-1}(w)}}.
\end{align}
If $A_1 > 0$, $W(t,f)$ is monotonic and the inverse has a single branch, this leads to a smooth work distribution. However, if $A_1 < 0$, $W(t,f)$ can develop three branches, with extrema at
\begin{align}\label{f12}
	f_{1,2} = \beta \sigma^2 \pm \frac{\sigma^2}{\mu} \cosh^{-1} \sqrt{-\frac{A_1(t) \mu}{\sigma^2}},
\end{align}
which exist for
\begin{align}
	\sigma < \sigma_c = \sqrt{-A_1(t) \mu}.
\end{align}
At the corresponding work values 
\begin{align}
	w_{1,2}=W(t,f_{1,2}) 
\end{align}
the distribution $P_W(w)$ can show peaks if $P_F(f_{1,2})$ is appreciable. The positions of $f_{1,2}$ are centered around $f^*=\beta\sigma^2$ and are symmetric with respect to it. Since $P_F(f)$ is appreciable only in a window around the eigenvalues of $\hat H(0)$ with a width of order $\sigma$, the visibility of the peaks depends on the interplay between $\sigma$, $A_1$, and the ratio of populations of the initial state, to which in a two-level system we can assign a temperature $T$. It is clear from (\ref{f12}) that if $f^*$ has to lie within an interval containing the eigenvalues of $\hat H(0)$, as the absolute value of $T$ decreases, $f_{1,2}$ will move outside the mentioned interval; this establishes a minimal $|T_c|$ below which one or two non-analyticities may appear in the work distribution. Notice that the non-analyticities appear also in the case of population inversion (negative temperatures). For $\sigma > \sigma_c$, $W(t,f)$ recovers its monotonic dependence on $f$ and $P_W(w)$ is again smooth.

We thus identify a \emph{resolution-driven transition} in $P_W(w)$ controlled by $A_1(t)$ and $\sigma$: the distribution is smooth for $\sigma>\sigma_c$, while for $\sigma<\sigma_c$, $A_1<0$, and $|T|\ge |T_c|$ it develops non-analytic peak structures. This is our second result. 

\subsection{Behavior of the peak separation near the transition and singular structure of $P_W(w)$}

Having identified the transition and on which parameters depends, we now analyse two quantitative results that characterize the work distribution and its non-analitic structure and which are directly experimentally accessible. Notice that non-analytic peaks are visible when we are in the intermediate resolution region ($\mu \approx \sigma$).

\paragraph{Scaling of the peak separation.}
The separation between the two singular work values,
\begin{equation}
	\Delta W = w_2 - w_1, \,\, (w_2 \ge w_1)
	\label{eq:DeltaW}
\end{equation}
constitutes a natural parameter to characterize the transition:
it vanishes identically for $\sigma\geq\sigma_c$ and is
positive whenever there are non-analyticities in the work distribution. Its exact expression follows from Eqs.~(\ref{eq:w_tanh_short}) and (\ref{f12}).
Evaluating $W(t,f)$ at the extrema, where
$\tanh(x)=\pm\sqrt{1-\sigma^2/\sigma_c^2}$ with
$x=\cosh^{-1}(\sigma_c/\sigma)$, one obtains
\begin{equation}
	\Delta W
	= \frac{2\sigma_c^2}{\mu}
	\sqrt{1-\frac{\sigma^2}{\sigma_c^2}}
	- \frac{2\sigma^2}{\mu}
	\cosh^{-1}\!\!\left(\frac{\sigma_c}{\sigma}\right).
	\label{eq:Wexact}
\end{equation}
Defining $\varepsilon = (\sigma_c-\sigma)/\sigma_c \ll 1$ to examine the behavior near the transition, each of the
two terms in~(\ref{eq:Wexact}) is individually of order $\varepsilon^{1/2}$ near $\sigma_c$.  However, their leading-order contributions \emph{cancel exactly}.
Expanding using
\begin{align}
	\sqrt{1-(1-\varepsilon)^2}
	&= \sqrt{2\varepsilon}\left(1-\frac{\varepsilon}{4}
	+O(\varepsilon^2)\right),
	\label{eq:sqrt_expand}\\
	\cosh^{-1}\!\!\left(\frac{1}{1-\varepsilon}\right)
	&= \sqrt{2\varepsilon}\left(1+\frac{5\varepsilon}{12}
	+O(\varepsilon^2)\right),
	\label{eq:acosh_expand}
\end{align}
the two $\varepsilon^{1/2}$ terms cancel and the leading
surviving contribution is of order $\varepsilon^{3/2}$,
\begin{equation}\label{eq:Wscaling}
\Delta W= \frac{8\sqrt{2}\,\sigma_c^2}{3\mu}\,\varepsilon^{3/2}
\left[1 - \frac{7}{20} \varepsilon + O(\varepsilon^2)\right].
\end{equation}
The exponent $3/2$ is universal: it does not depend on the details of the Hamiltonian, the temperature, the
exact form of $P_F(f)$, or $A_1(t)$. It follows solely from the fact that the two extrema of $W(t,f)$ coalesce
at $\sigma=\sigma_c$, which forces $W(t,f)$ to have a cubic inflection point there. Near such a point the extrema separate as $\Delta f \sim \varepsilon^{1/2}$, and since $W$ is locally cubic, the corresponding work separation scales as $\Delta W \sim (\Delta f)^3 \sim \varepsilon^{3/2}$, independently of any microscopic detail.


\begin{figure}[t]
	\centering
	\includegraphics[width=.9\linewidth]{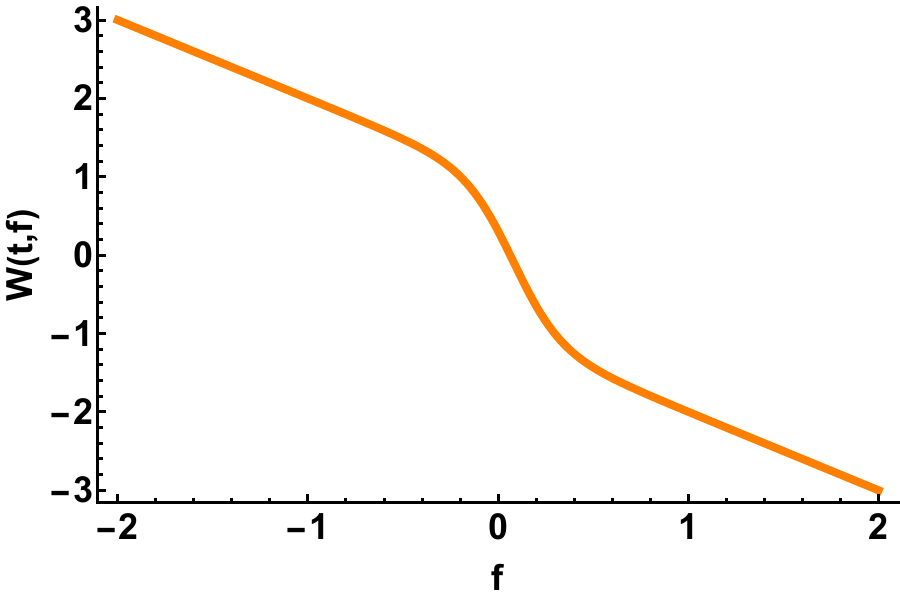}\\[2mm]
	\includegraphics[width=.9\linewidth]{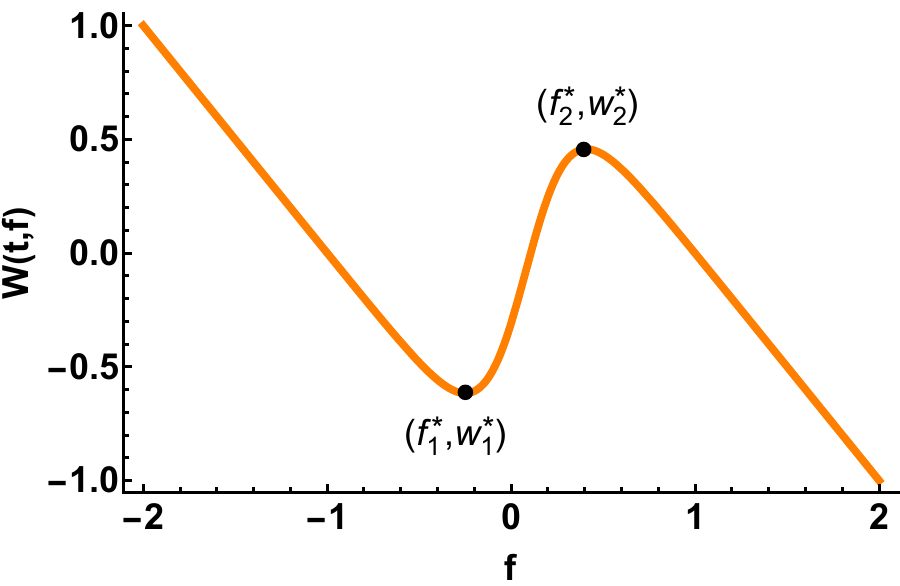}
	\caption{Function $W(t,f)$ for a given final time $t$ with positive (top) and negative (bottom) $A_1$. For $A_1(t)>0$ the inverse $f=W^{-1}(w)$ has a single branch and yields a smooth $P_W(w)$. For $A_1(t)<0$ three branches appear and there are work values $w^*_{1,2}$ for which $dW/df=0$. If $P_F(f^*_{1,2})$ is non-negligible, $P_W(w)$ develops peaks at $w^*_{1,2}$.}
	\label{fig:figure1}
\end{figure}

\paragraph{Singular structure of $P_W(w)$ near
	$w_{1,2}$.}
At each singular work values $w_i$ ($i=1,2$), the distribution $P_W(w)$ exhibits a characteristic
square-root divergence. Expanding $W(t,f)$ to second order around the extremum $f_i$,
\begin{equation}
	W(t,f)-w_i \approx \tfrac{1}{2}W''(f_i)(f-f_i)^2,
\end{equation}
and noting that the second derivative at the extrema vanishes as 
\begin{equation}
	|W''(f_{1,2})| \approx \frac{2\mu^2}{\sigma_c^3}
	\sqrt{2(\sigma_c-\sigma)} \sim \varepsilon^{1/2},
	\label{eq:W2nd}
\end{equation}
the Jacobian of the change of variables vanishes as $|dW/df|\approx\sqrt{2|W''(f_i)|\,|w-w_i|}$, and
substituting into Eq.~(\ref{eq:PWbranches}) yields
\begin{equation}
	P_W(w) \underset{w\to w_i}{\approx}
	\frac{P_F(f_i)}{\sqrt{2\,|W''(f_i)|\,|w-w_i|}},
	\quad i=1,2.
	\label{eq:PWsing}
\end{equation}
This $|w-w_i|^{-1/2}$ divergence is integrable, the integral of $P_W(w)$ over any neighbourhood of $w_i$
converges, but produces a pronounced peak in the histogram of work values. The amplitude is controlled by two factors: $P_F(f_i)$ and $|W''(f_i)|^{-1/2}$, which diverges as $(\sigma_c-\sigma)^{-1/4}$ from (\ref{eq:W2nd}). The peak height therefore grows as
\begin{equation}
	P_W(w_i) \sim P_F(f_i)\,\frac{\sigma_c^{3/2}}{2\mu}
	\left(\sigma_c-\sigma\right)^{-1/4}|w-w_i|^{-1/2},
	\label{eq:PWsingfull}
\end{equation}
providing a second, independently measurable signature of the critical point. Equations~(\ref{eq:Wscaling}) and~(\ref{eq:PWsing})--(\ref{eq:PWsingfull}) constitute
the quantitative content of this section: a complete characterization of the non-analytic structure of $P_W(w)$ near the transition, with two independently measurable signatures accessible in current NV-center experiments.

\section{Extracting coherence from the OTM work distribution}\label{coherences}

The following section shows how quantum coherences can be inferred from the OTM work distribution in a two-level system.

Starting from the initial state (\ref{inistate}), the conditional state after an outcome $f$ can be expressed in the eigenbasis of $\hat H(0)$ as
\begin{eqnarray}
	\rho(t|f) = P_F^{-1}(f) \sum_{i=1,2} &p(\mu_i(0),\beta) G_\sigma(f|\mu_i(0)) \nonumber \\ &\times \hat U(t)|\mu_i(0)\rangle\langle \mu_i(0)|\hat U^\dagger(t),
\end{eqnarray}
with $P_F(f) = \sum_{i=1,2} p(\mu_i(0),\beta) G_\sigma(f|\mu_i(0))$. The unconditional state
\begin{equation}
	\rho(t) = \sum_{i=1,2} p(\mu_i(0),\beta) \hat U(t) |\mu_i(0)\rangle\langle \mu_i(0)| \hat U^\dagger(t),
\end{equation}
follows when averaging over the outcomes $f$. This state generally exhibits coherence in the instantaneous eigenbasis of $\hat H(t)$.

The populations in this basis are
\begin{equation}\label{populations}
	\rho_{kk}(t) = \sum_{i=1,2} p(\mu_i(0),\beta), |\langle \mu_k(t) | \hat U(t) | \mu_i(0) \rangle|^2.
\end{equation}
The relative entropy of coherence in the $\hat H(t)$ basis is
\begin{eqnarray}
	C_{\hat H(t)} &=& - \sum_{k=1,2} \rho_{kk}(t) \ln \rho_{kk}(t) \nonumber \\
	 &+& \sum_{i=1,2} p(\mu_i(0),\beta) \ln p(\mu_i(0),\beta),
\end{eqnarray}
which quantifies the coherence resource in the final state \cite{streltsov2017quantum}.

For a two-level system the populations \eqref{populations} can be expressed in terms of $\Delta p = p(\mu_2(0),\beta) - p(\mu_1(0),\beta)$ and $A_1(t)$ as
\begin{align}
	\rho_{11}(t) &= \frac{1}{2}\left[1 + \Delta p, \frac{A_1(t)}{\mu_2(t)}\right], \
	\rho_{22}(t) &= \frac{1}{2}\left[1 - \Delta p, \frac{A_1(t)}{\mu_2(t)}\right],
\end{align}
where $\mu_2(t)$ is the largest eigenvalue of $\hat H(t)$. Thus, once the combinations $A_1(t)\Delta p$ and $\mu_2(t)$ are determined from the work distribution, the populations and hence $C_{\hat H(t)}$ are fully specified.

The same dynamical quantities that govern the appearance of non-analyticities in $P_W(w)$ (such as the critical resolution $\sigma_c$ and the positions of the peaks at $w_{1,2}$) therefore encode the coherence content of the final state. In this way, the OTM work distribution provides a direct and experimentally feasible route to quantify coherence. This is our third main result.

\section{Experimental realization: NV centers in diamond}\label{Exp}

We now discuss an experimentally relevant implementation of our ideas in the NV-center platform used in \cite{hernandez2020experimental} to test quantum fluctuation relations in an open system. There, the electronic spin of the NV center is driven by microwave (GHz) and radiofrequency (MHz) pulses and coupled to a surrounding nuclear spin bath. The experiment demonstrates exchange fluctuation relations for energy and spin in a non-equilibrium setting, confirming that fluctuation theorems hold in open quantum systems with coherent dynamics. The NV center, with its high degree of control and long coherence times, is an ideal platform for probing quantum thermodynamic relations, including work statistics and coherence effects.

To connect with this platform, we model the NV electronic spin as a two-level system driven by a microwave field,
\begin{equation}
	\hat H(t) = \frac{\hbar \omega_0}{2} \hat \sigma_z + \frac{\hbar \omega_1}{2} \big( \hat \sigma_x \cos \omega t + \hat \sigma_y \sin \omega t \big),
\end{equation}
with Pauli matrices $\hat \sigma_{x,y,z}$. Moving to the rotating frame defined by $\hat U_1 = \exp(-i \omega t \hat \sigma_z / 2)$, the dynamics is governed by the time-independent Hamiltonian
\begin{equation}
	\hat{\bar H} = \frac{\hbar \delta}{2} \hat \sigma_z + \frac{\hbar \omega_1}{2} \hat \sigma_x,
\end{equation}
where $\delta = \omega_0 - \omega$ is the detuning. This describes a standard Rabi model with tunable detuning and drive amplitude \cite{ribeiro2016quantum}. 

The dynamical factor $A_1(t)$ can be computed analytically. In particular, one finds
\begin{eqnarray}
	A_1(t)/\hbar=\frac{-\omega^2 \omega_1^2 \cos(\Omega t) - (\omega_0^2 - \omega \omega_0 + \omega_1^2)^2}{2 \Omega^2 \sqrt{\omega_0^2 + \omega_1^2}},
\end{eqnarray}
with $\Omega = \sqrt{\omega_1^2 + \delta^2}$. The sign of $A_1(t)$ is determined by
\begin{equation}\label{eq:A1sign}
	s = -\big[ (\delta + \omega)^2 - \omega (\delta + \omega) + \omega_1^2 \big]^2 - \omega^2 \omega_1^2 \cos(\Omega t),
\end{equation}
which defines regions in the $(t,\delta)$ plane where $A_1(t)$ is positive or negative. For $0 < \delta < \delta_c = \frac{1}{2} \sqrt{\omega^2 + 4 \omega_1 \omega - 4 \omega_1^2} - \frac{\omega}{2}$, $A_1(t)$ can change sign, while for $\delta > \delta_c$ it is strictly negative. Under typical experimental conditions, with $\omega \gg \omega_1$, one has $\delta_c \simeq \omega_1 (1 - 2 \omega_1 / \omega + \dots)$.

\begin{figure}[t]
	\centering
	\includegraphics[width=.8\linewidth]{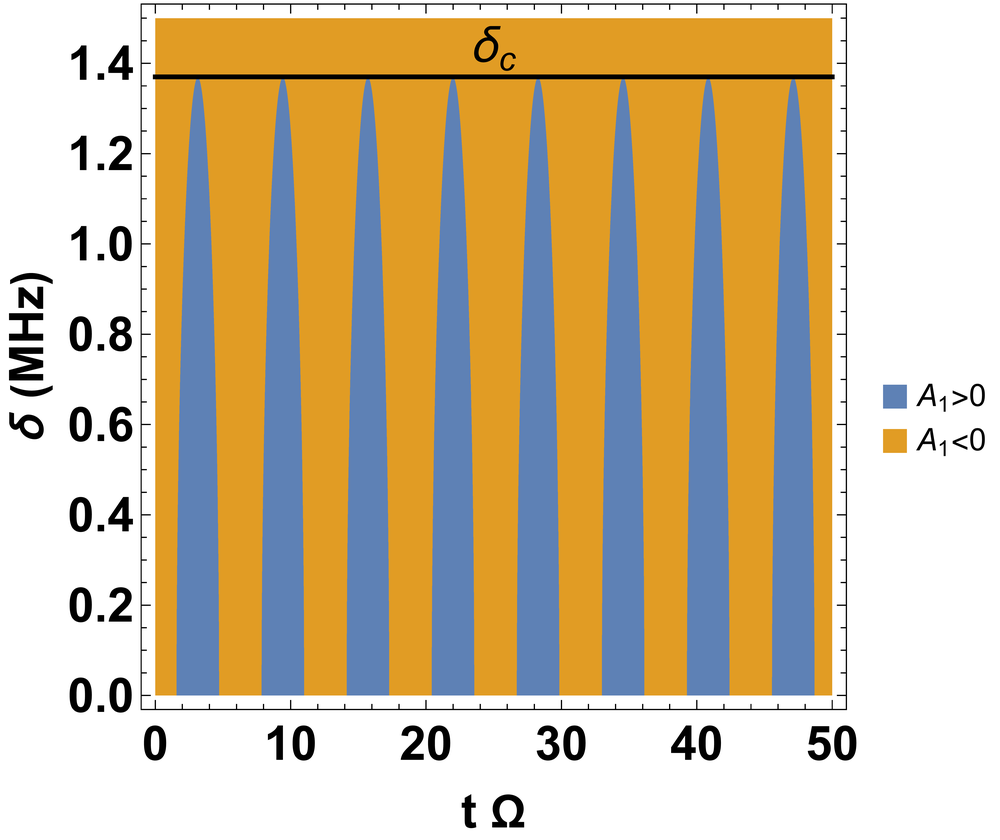}
	\caption{Sign of $A_1$ as a function of final time $t$ and detuning $\delta$. The horizontal dashed line marks the critical detuning $\delta_c$. Parameters as in Ref.~\cite{hernandez2020experimental}: $\omega=3.0 \times 10^9\,\text{Hz}$ and $\omega_1 = 1.37 \times 10^6\,\text{Hz}$.}
	\label{fig:figure2}
\end{figure}

Fig.(\ref{fig:figure2}) illustrates the regions in the $(t,\delta)$ plane where $A_1$ is positive or negative for experimental parameters taken from Ref.~\cite{hernandez2020experimental}. Regions with $A_1>0$ correspond to smooth OTM work distributions, while regions with $A_1<0$ can exhibit the non-analytic peak structure associated with the statistical transition described above. The NV-center platform thus provides direct access to both sides of the transition.

\section{Numerical work distributions}\label{num}

To further illustrate the statistical transition, we perform numerical simulations of the work distribution $P_W(w)$ for the driven NV-center Hamiltonian. 

For a given set of parameters $(\omega,\omega_1,\delta)$ and a fixed final time $t$, we first compute the probability density of energy-measurement outcomes, $P_F(f)$, using Eq. \eqref{eq:work} and the Gaussian meter model. We then generate a large ensemble of outcomes $f$ sampled from $P_F(f)$ and, for each realization, compute the corresponding work $W(t,f)$. The resulting ensemble of work values defines a numerical realization of $P_W(w)$.

For $A_1(t)>0$ and large $\lambda$, $P_W(w)$ displays a two-Gaussian structure corresponding to the two possible energy transitions. As the resolution is reduced (smaller $\lambda$), the peaks broaden and eventually merge into a single, smooth peak, approaching the weak-measurement regime.

\begin{figure*}[t]
	\centering
	(a){\includegraphics[width=0.4\textwidth]{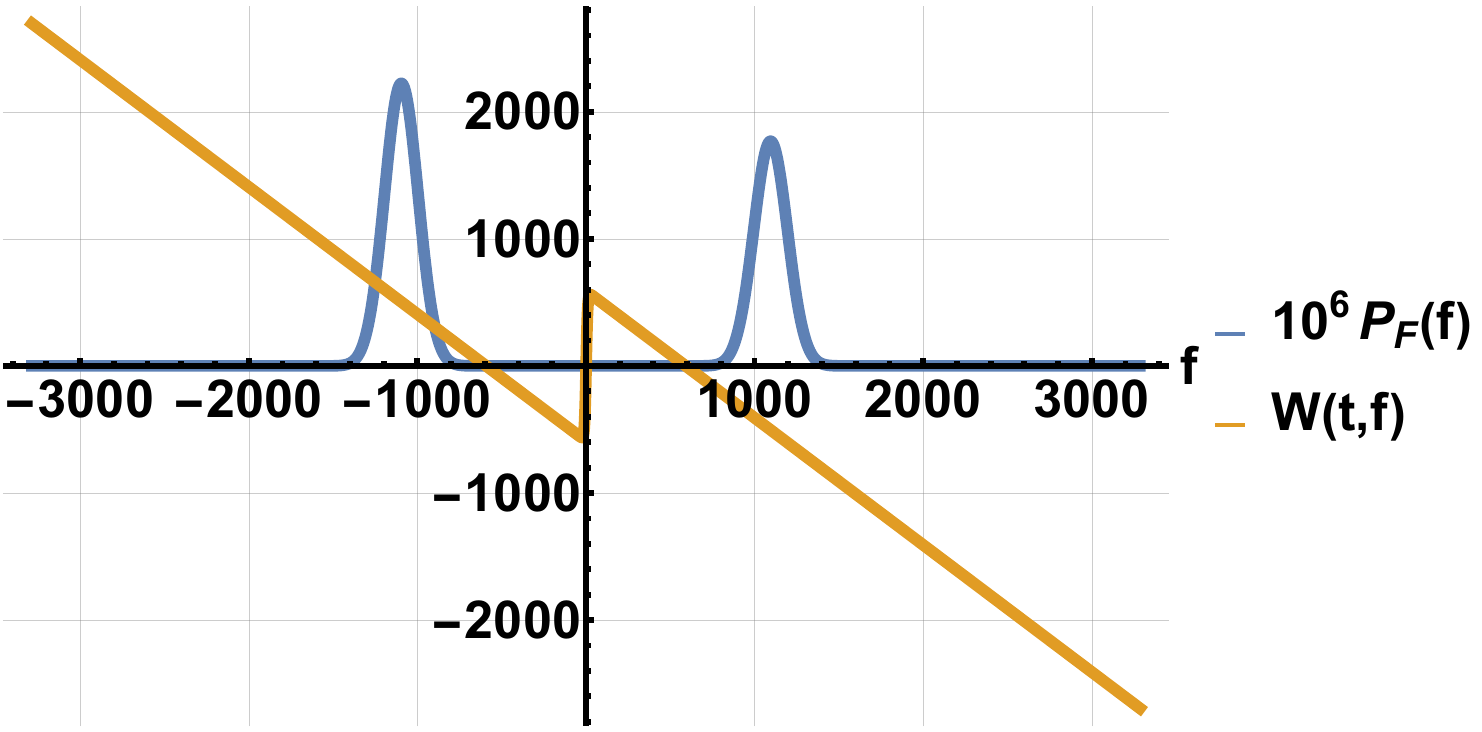}}
	(b){\includegraphics[width=0.4\textwidth]{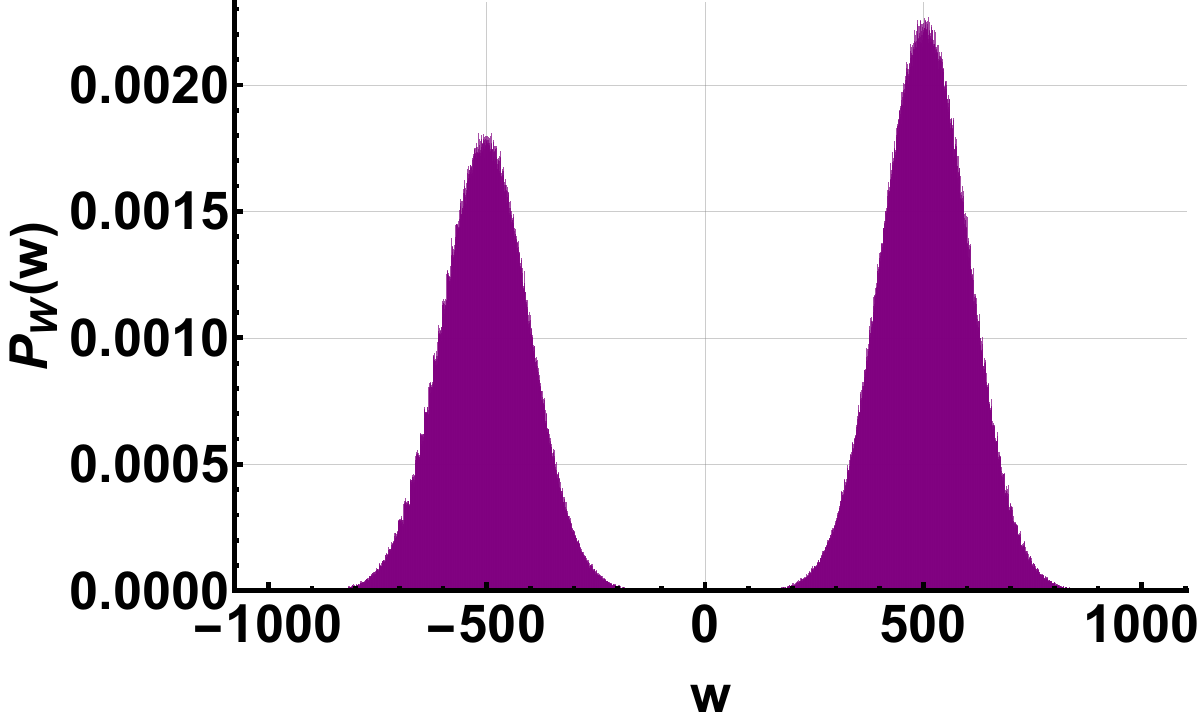}}
	(c){\includegraphics[width=0.4\textwidth]{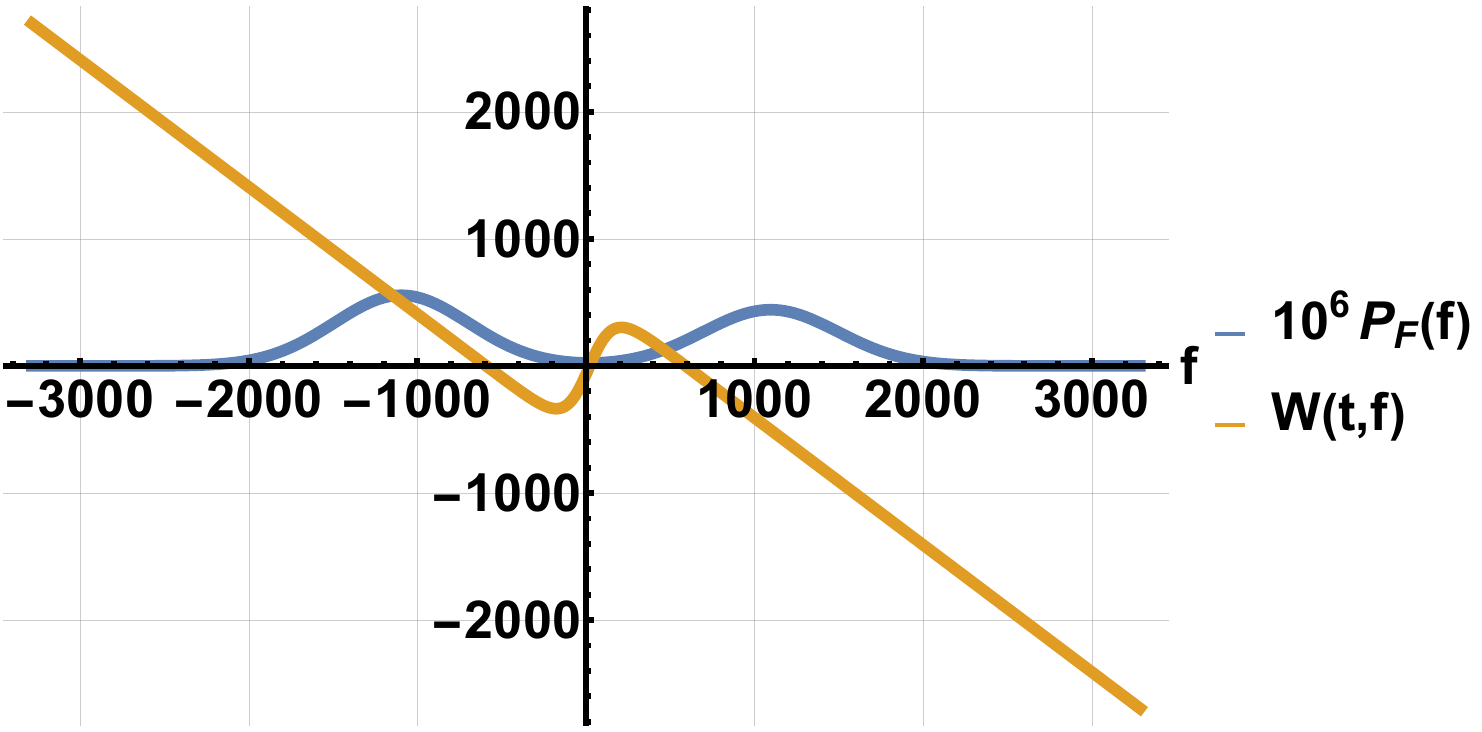}}
	(d){\includegraphics[width=0.4\textwidth]{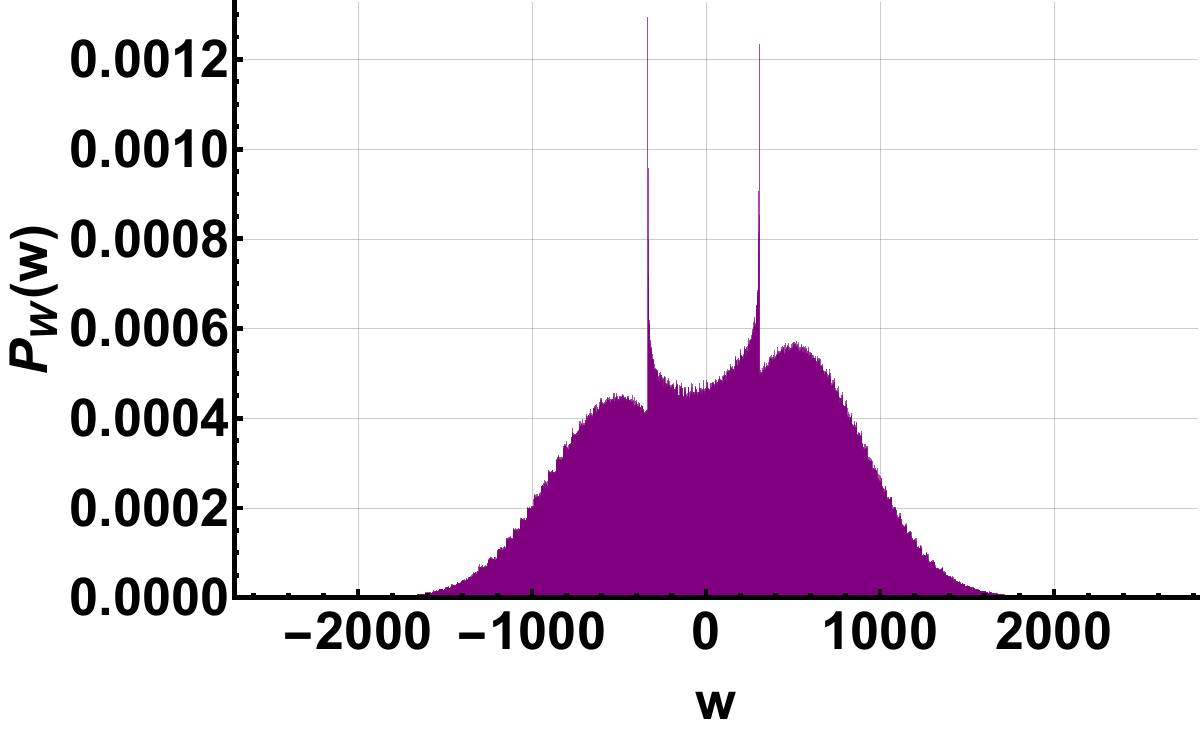}}
	(e){\includegraphics[width=0.4\textwidth]{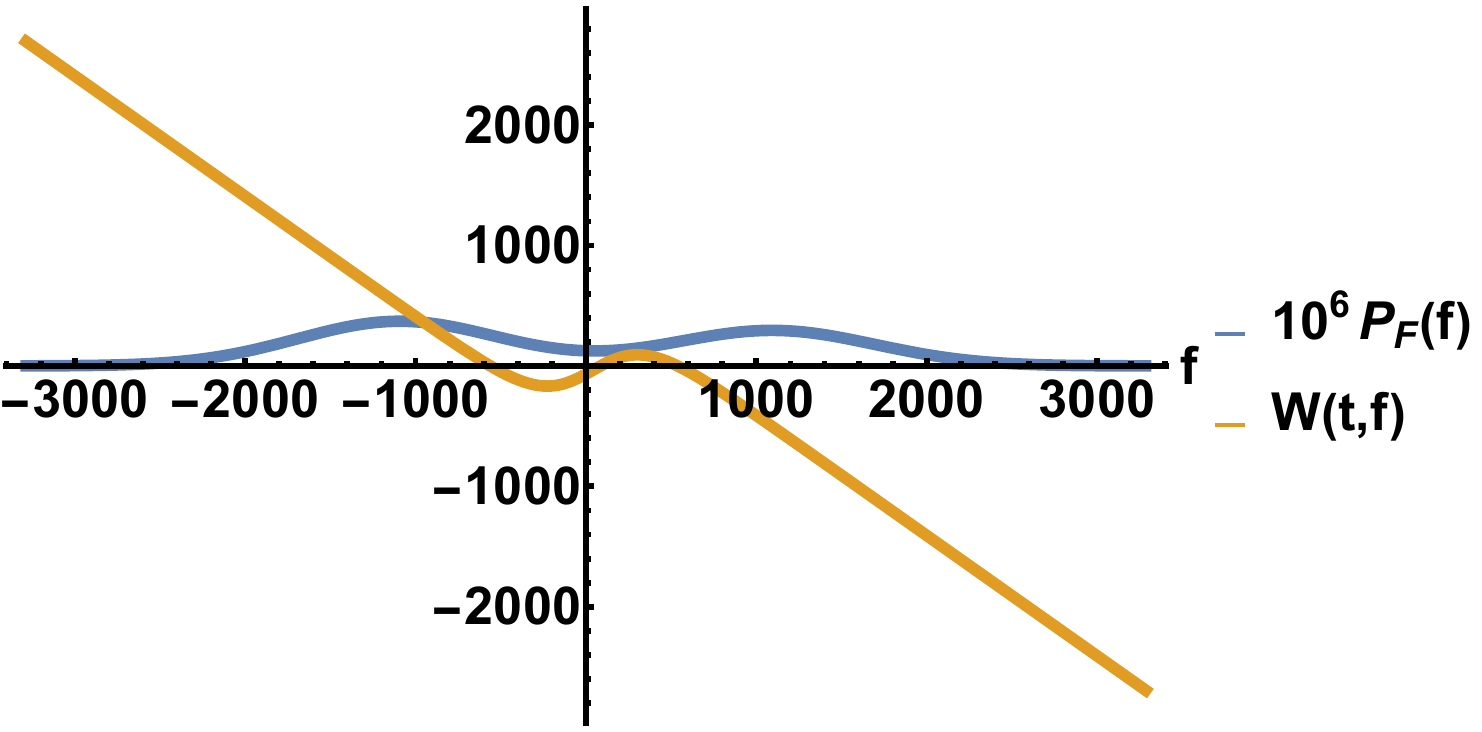}}
	(f){\includegraphics[width=0.4\textwidth]{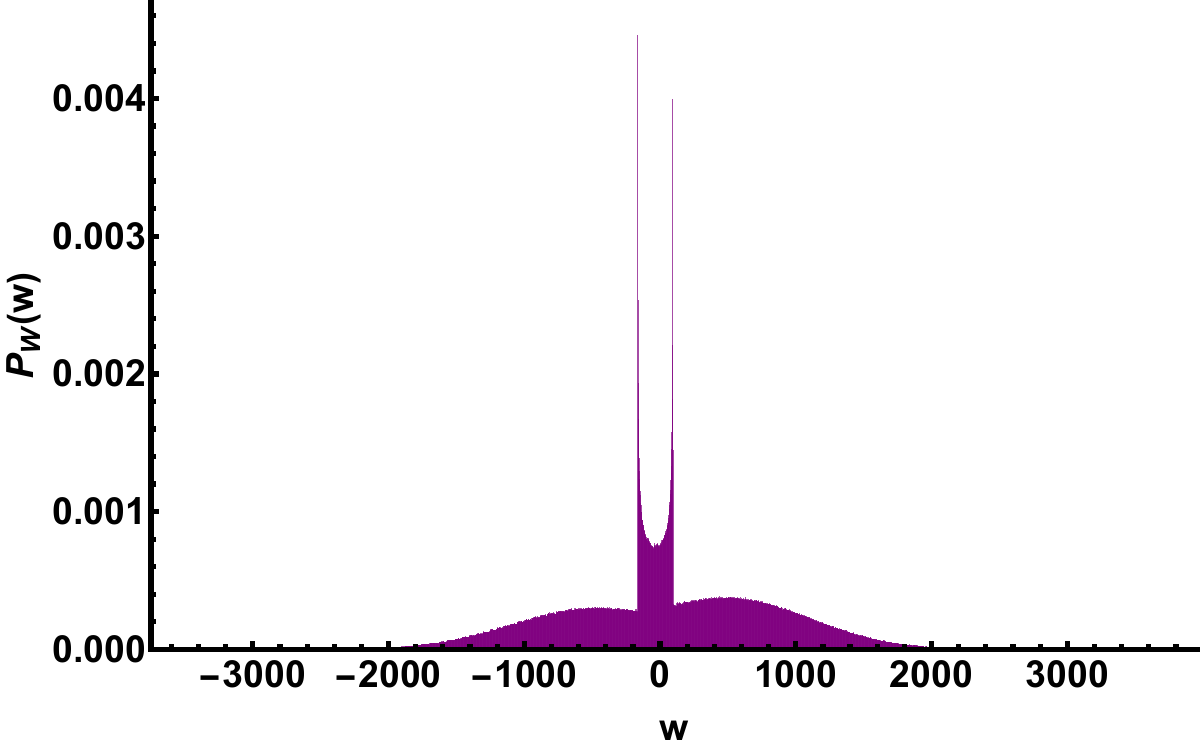}}
	\caption{Sequence of work probability distributions for decreasing values of $\lambda$ and $A_1<0$. For each value of $\lambda$, the left panel shows $P_F(f)$ and $W(t,f)$, while the right panel depicts the resulting $P_W(w)$. Large $\lambda$ corresponds to ideal measurements, while smaller $\lambda$ approaches the weak-measurement limit. The model parameters are: $\omega=3 \times 10^9$ Hz, $\omega_1 = 1.37 \times 10^6$ Hz, $\delta = 0$, and $t = 1 \times \, \Omega^{-1}$, yielding $A_1<0$. In panels (a)-(b): $\lambda = 21.9$, (c)-(d): $\lambda = 5.47$, (e)-(f): $\lambda = 3.65$. Energies $f, w,$ and $W(t,f)$ are expressed in units of $\hbar \Omega$.}
	\label{fig:3}
\end{figure*}

\begin{figure*}[t]
	\centering
	(a){\includegraphics[width=0.4\textwidth]{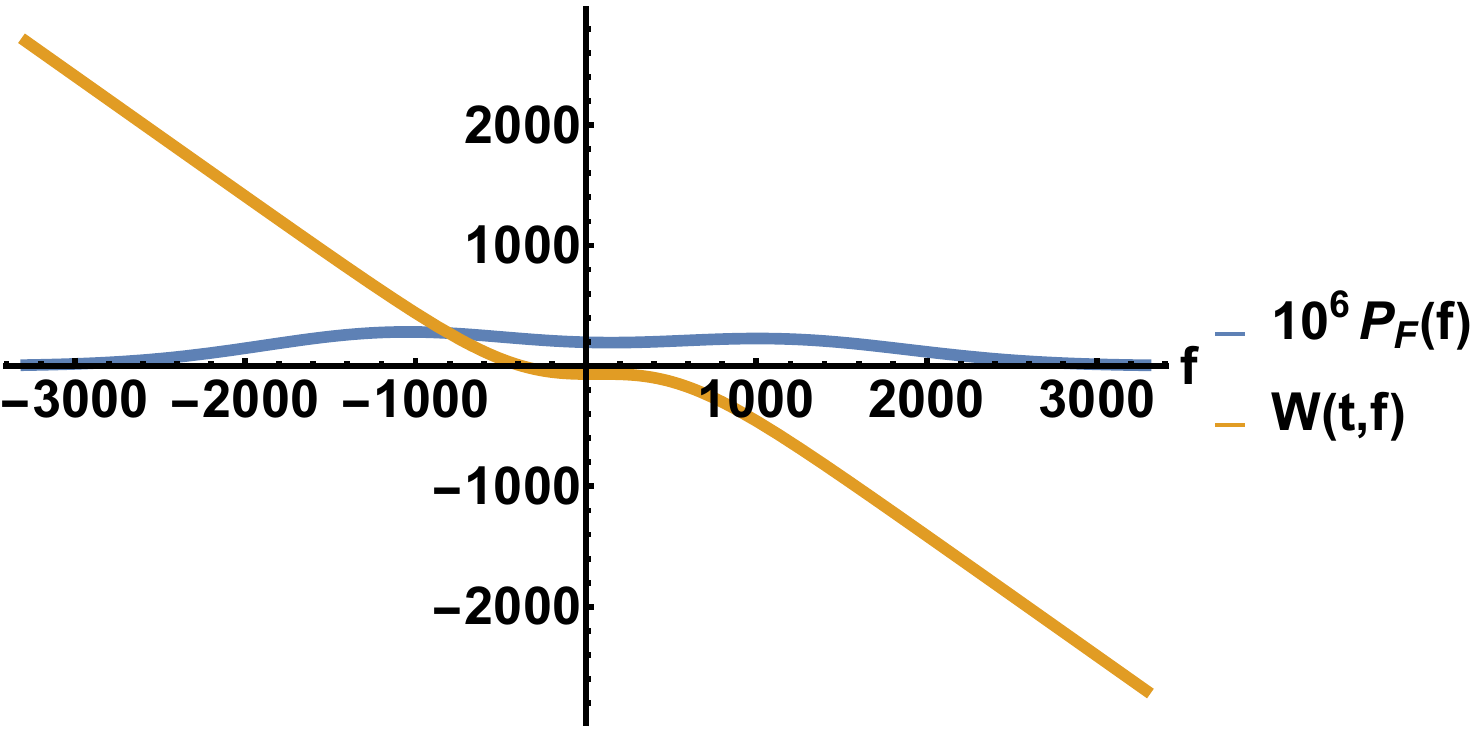}}
	(b){\includegraphics[width=0.4\textwidth]{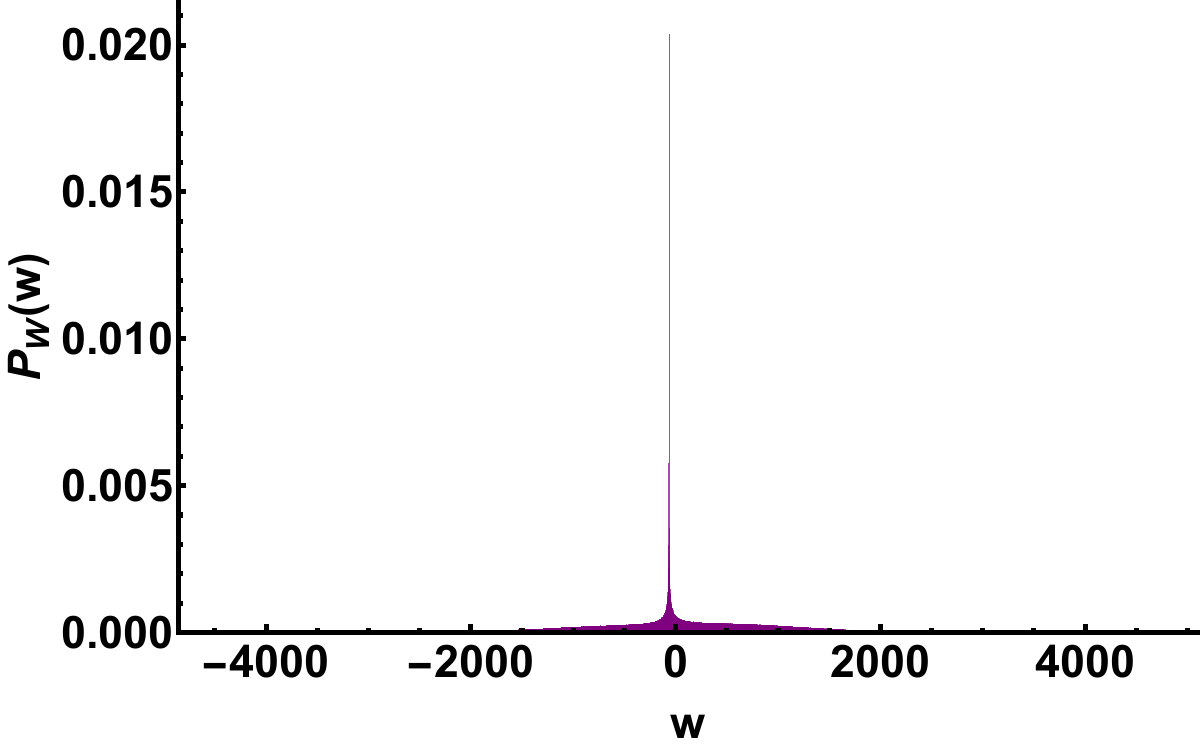}}
	(c){\includegraphics[width=0.4\textwidth]{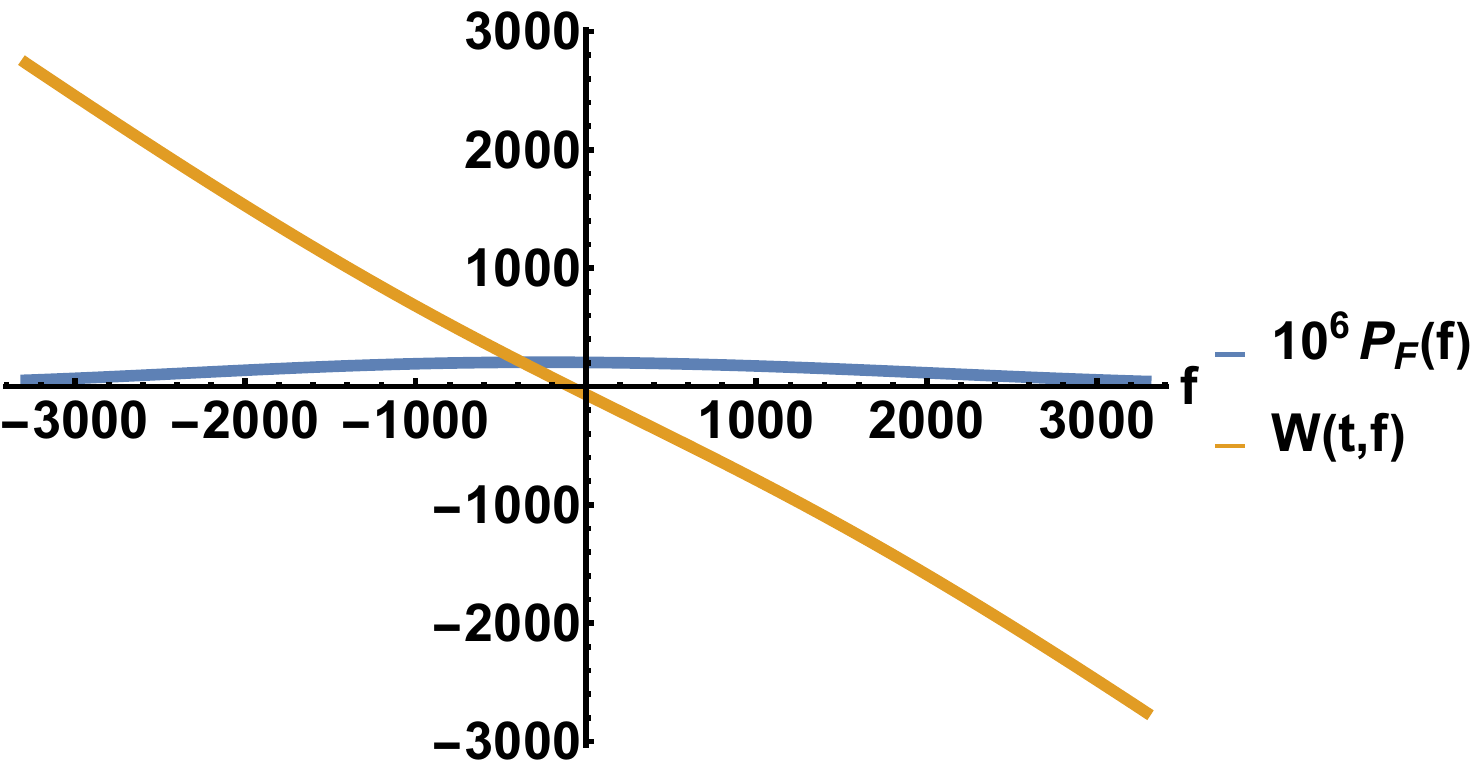}}
	(d){\includegraphics[width=0.4\textwidth]{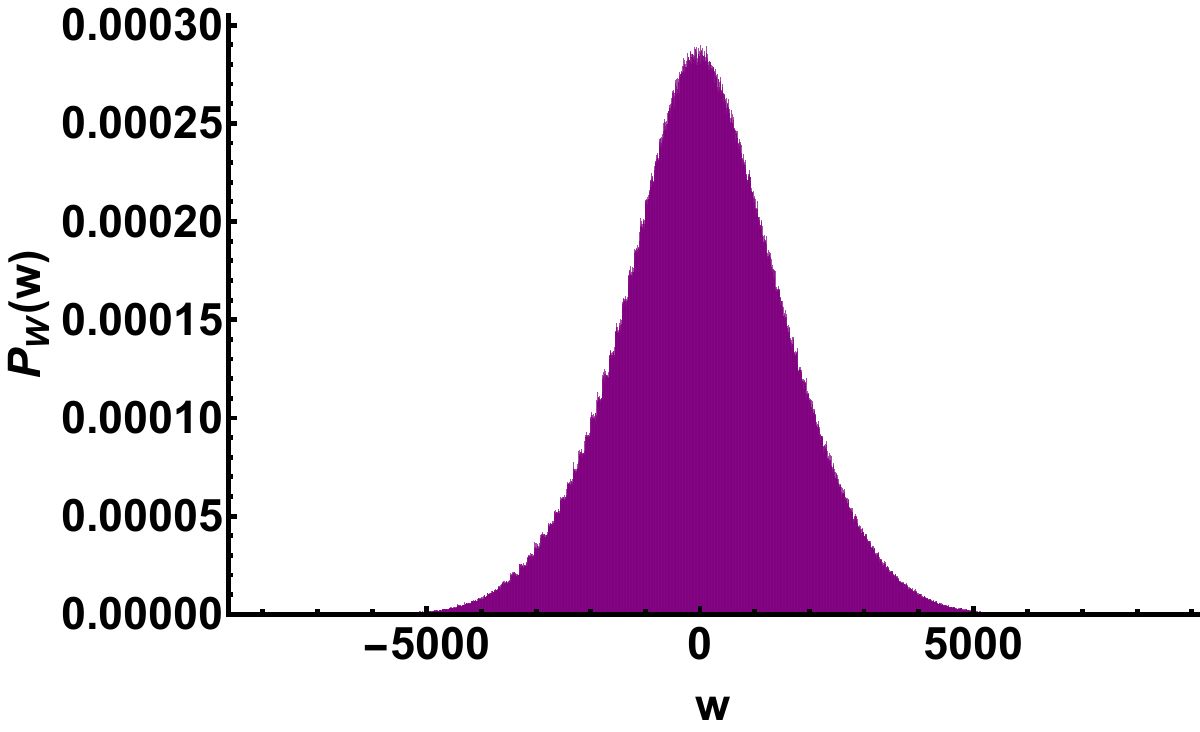}}
	\caption{The same as Fig. \ref{fig:3}. In panels (a)-(b): $\lambda = 2.72$, which corresponds to $\omega_1^* = \omega_2^*$, and in panels (c)-(d) $\lambda = 1.46$. This last value approaches the weak-measurement regime.}
	\label{fig:4}
\end{figure*}

In Fig.(\ref{fig:3}) and Fig.(\ref{fig:4}) we depict $P_W(w)$ for the different situations discussed in the case $A_1(t)<0$ and large $\lambda$, $P_W(w)$ again shows two well-separated peaks. As $\sigma$ is increased towards $\sigma_c$, additional sharp features emerge at work values $w_{1,2}$ associated with the extrema of $W(t,f)$. Near $\sigma_c$, the peaks merge into a single pronounced peak. For $\sigma>\sigma_c$, the distribution gradually gets smoother and tends to a Gaussian-like shape at very low resolution.

This behavior confirms the analytical picture developed in section \ref{TLS}: the sign of $A_1(t)$ and the value of $\lambda$ determine whether the work distribution is smooth or exhibits the non-analytic peak structure associated with the transition. 

\begin{figure*}[t]
	\centering
	{\includegraphics[width=0.6\textwidth]{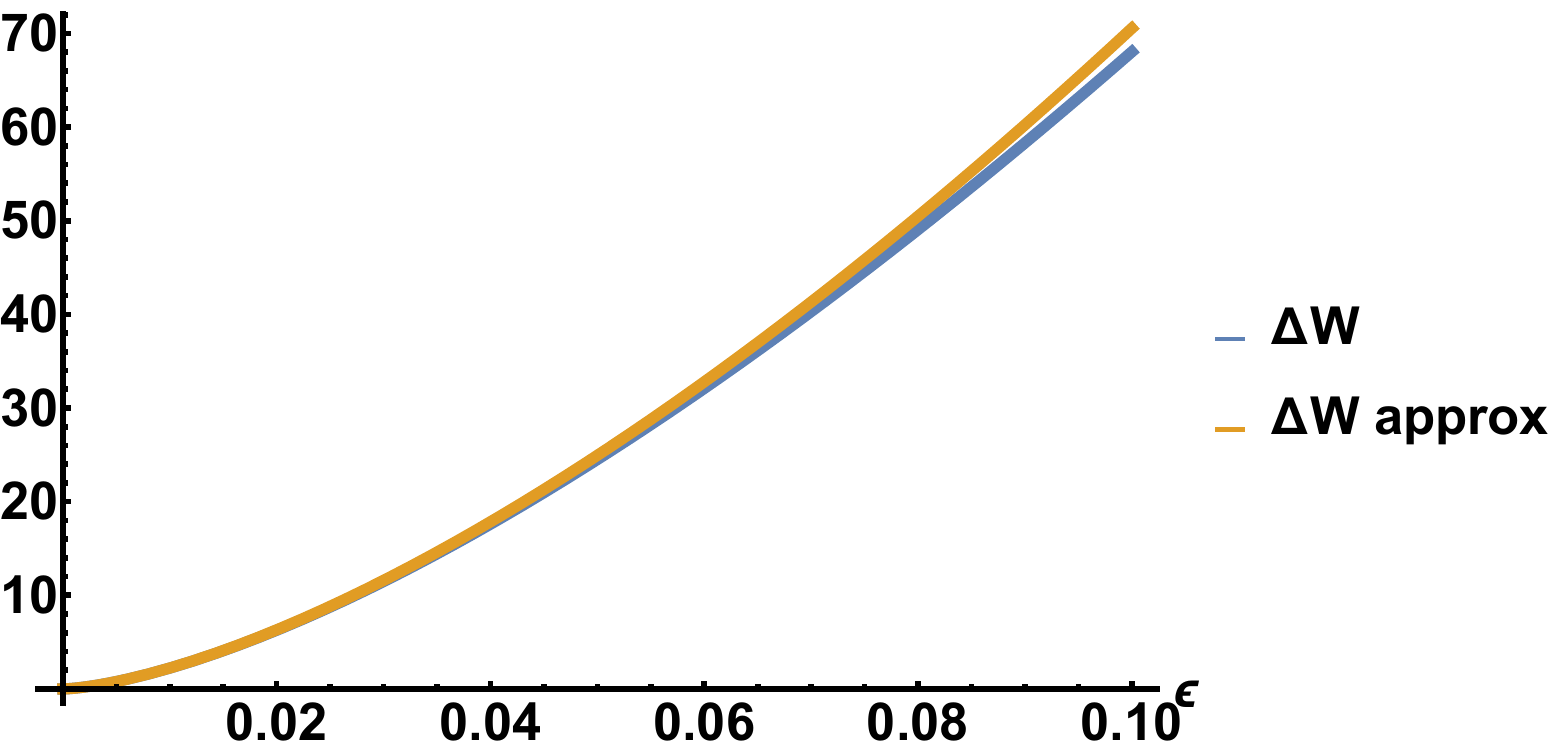}}
	\caption{Peak separation $\Delta W = w_2 - w_1$ in the NV-center system as a function of the scaling parameter
		$\varepsilon = (\sigma_c-\sigma)/\sigma_c$. The exact expression, Eq.~(\ref{eq:Wexact}), is compared to the
		approximated formula, $\Delta W(\sigma) = \frac{8\sqrt{2}\,\sigma_c^2}{3\mu}\,
		\varepsilon^{3/2}\!\left[1-\frac{7}{20}\varepsilon\right]$, confirming the universal scaling exponent $3/2$. Parameters
		as in Ref.~[11]. $\Delta W$ is expressed in units of $\hbar\Omega$.}
	\label{fig:5}
\end{figure*}

Fig.(\ref{fig:5}) shows how the peak separation $\Delta W$ scales near the transition as a function of the parameter
$\varepsilon = (\sigma_c - \sigma)/\sigma_c$, for the NV-center parameters of Ref. \cite{hernandez2020experimental}. The agreement between the exact expression (\ref{eq:Wexact}) and the approximate formula (\ref{eq:Wscaling}) is excellent for $\varepsilon \lesssim 0.08$, this gives a set of parameters for which both behaviors, analytic and non-analytic, can be observed for realistic meter resolutions and population ratios accessible with current technology.

\section{Conclusions}

We have shown that OTM work statistics can be reconstructed from standard TTM data through classical post-processing. In this manner, we establish an explicit operational connection between these two widely used protocols. This finding indicates that TTM data can encode coherence-related information that is commonly associated with OTM schemes, and that such information may be accessed without modifying the experimental setup.

For the case of a driven two-level system subject to finite resolution energy measurements, we identified a resolution-induced non-analytic change in the structure of the OTM work distribution, which we refer to as a \emph{statistical transition}, governed by the sign of a dynamical quantity $A_1$ and a critical meter resolution $\sigma_c$. Whereas projective and weak measurements yield smooth, Gaussian-like distributions, intermediate resolutions can result in non-analytic peaks associated with multiple branches of the conditional work function $W(t,f)$. We characterize these non-analyticities quantitatively: the separation between the two singular work values scales as $\Delta W \sim \varepsilon^{3/2}$ near the critical resolution, where $\varepsilon = (\sigma_c-\sigma)/\sigma_c$, with a universal exponent arising from the coalescence of the two critical points of $W(t,f)$ at $\sigma_c$. The work distribution diverges at $w_{1,2}$ as  $P_W(w) \sim 	\left(\sigma_c-\sigma\right)^{-1/4} |w-w_i|^{-1/2}$ as $w \to w_i$.

Finally, we showed that, for two-level systems, the OTM work distribution contains sufficient information to reconstruct the relative entropy of coherence at the final time. This provides a direct approach to coherence metrology based exclusively on energy measurements. To demonstrate the possibility of experimentally observing our results, we also discussed an NV-center implementation that should be observable with current technology.

Our results highlight the coherence-related information contained in standard TTM protocols and suggest that hybrid measurement and post-processing strategies are possible means to probe coherence in quantum thermodynamic processes. Extensions to open systems, applications in quantum sensing, and information-driven thermodynamics represent promising directions for future research.

\section*{Acknowledgements}
D.A. expresses his deepest gratitude to Professor Pierre Gaspard for his invaluable mentorship, continuous encouragement, and inspiring guidance in scientific research over the years. We also thank, Gabriele De Chiara, Inés de Vega, Fernando Delgado, Felipe Barra, Antonio Alejandro Valido and Luis Correa for their helpful comments and suggestions. Financial support from the Spanish Ministry of Science and Innovation through project PID2022-138269NB-I00 (MINECO/FEDER, UE) is gratefully acknowledged.

\section{Data Availability}
The analytical derivations are presented in the manuscript. Numerical codes used to generate the figures are available from the corresponding author upon reasonable request. 

\bibliographystyle{apsrev4-2}

\end{document}